\documentclass[a4paper,fleqn]{cas-dc}

\usepackage[numbers]{natbib}
\usepackage{xcolor}
\usepackage{subfig}
\usepackage{bm}
\usepackage{xcolor}
\usepackage{amsmath}
\usepackage{cleveref}
\Crefname{equation}{Equation}{Eqs.}

\def\tsc#1{\csdef{#1}{\textsc{\lowercase{#1}}\xspace}}
\tsc{WGM}
\tsc{QE}

\begin{document}
\let\WriteBookmarks\relax
\def\floatpagepagefraction{1}
\def\textpagefraction{.001}

\shorttitle{Scenario analysis of livestock-related PM$_{2.5}$ pollution}    

\shortauthors{J. Rodeschini, A. Fassò, F. Finazzi and A. Fusta Moro}  

\title [mode = title]{Scenario analysis of livestock-related PM$_{2.5}$ pollution based on a new heteroskedastic spatiotemporal model} 

\author[1]{Jacopo Rodeschini}[orcid=0000-0003-1874-1854]
\cormark[1]

\ead{jacopo.rodeschini@unibg.it}



\author[2]{Alessandro Fassò}[orcid=0000-0001-5132-9488]
\ead{alessandro.fasso@unibg.it}

\author[2]{Francesco Finazzi}[orcid=0000-0002-1295-7657]
\ead{francesco.finazzi@unibg.it}

\author[1]{Alessandro {Fusta Moro}}[orcid=0000-0003-1129-5038]
\ead{alessandro.fustamoro@unibg.it}

\affiliation[1]{organization={Department of Engineering and Applied Science (DISA), University of Bergamo},
  addressline={Via Pasubio 3}, 
 city={Dalmine (BG)},
            postcode={24044}, 
            state={Italy},
            }

\affiliation[2]{organization={Department of Economics, University of Bergamo},
  addressline={Via dei Caniana 2}, 
            city={Bergamo},
            postcode={24127}, 
            state={Italy},
            }

\cortext[1]{Corresponding author}

\begin{abstract}
The air in the Lombardy Plain, Italy, is one of the most polluted in Europe due to limited atmosphere circulation and high emission levels. There is broad scientific consensus that ammonia (NH$_3$) emissions have a primary impact on air quality, and, in Lombardy, the agricultural sector and livestock activities are widely recognised as being responsible for approximately 97\% of regional ammonia emissions due to the high density of livestock. 

In this paper, we quantify the relationship between ammonia emissions and PM$_{2.5}$ concentrations in the Lombardy Plain and evaluate PM$_{2.5}$ changes due to the reduction of ammonia emissions through a "what-if" scenario analysis. The information in the data is exploited using a spatiotemporal statistical model capable of handling spatial and temporal correlation, as well as missing data. To do this, we propose a new heteroskedastic extension of the well-established Hidden Dynamic Geostatistical Model. Maximum likelihood parameter estimates are obtained by the expectation-maximisation algorithm and implemented in a new version of the D-STEM software.

Considering the years between 2016 and 2020, the scenario analysis is carried out on high-resolution PM$_{2.5}$ maps of the Lombardy Plain. As a result, it is shown that a 26\% reduction in NH$_3$ emissions in the wintertime could reduce the PM$_{2.5}$ average by $1.44$ $\mu g / m^3$ while a 50\% reduction could reduce the PM$_{2.5}$ average by $2.76$ $\mu g / m^3$ which corresponds to a reduction close to 3.6\% and 7\% respectively. Finally, results are detailed by province and land type. 
\end{abstract}






\maketitle

\section{Introduction}\label{Introduction}

Air pollutants can be categorised into two groups: primary and secondary. Primary pollutants are those directly emitted into the atmosphere, while secondary pollutants are formed in the atmosphere through chemical reactions and microphysical processes involving precursor gases. Ammonia (NH$_3$) is a key precursor gas for secondary particulate matter (PM), specifically PM$_{10}$ and PM$_{2.5}$. In the Lombardy region, the primary sources of ammonia are livestock and fertilisers, accounting for approximately 97\% of the overall regional emissions \cite{clappier2021impact, lonati2020temporal, lovarelli2021comparison}. Although total annual NH$_3$ emissions remain relatively constant over the considered period, variations within the year are influenced by seasonal agricultural practices. 
In the Lombardy Plain, due to limited air circulation and stability in the air, PM$_{2.5}$ frequently accumulates at hazardous concentrations, posing risks to human health. The elevated pollution levels in Lombardy Plain have led to one of the highest PM$_{2.5}$-related mortality rates in Europe, with 100-150 premature deaths per 100,000 inhabitants \cite{EEARisk}.


Using chemical transport model (CTM) simulations, several studies showed that within Europe, reducing NH$_3$ emissions stands out as one of the most efficient control strategies for mitigating PM$_{2.5}$ levels in both summer and winter seasons. In particular, \cite{megaritis2013response} used five distinct control strategies, with a particular focus on testing a 50\% reduction in gaseous emissions (SO$_2$, NH$_3$, NO$_x$ and anthropogenic volatile organic compounds) to assess concentration sensitivity to emissions. The findings indicated that, in the majority of European regions, reducing NH$_3$ emissions during winter and summer periods proves more efficacious in lowering overall PM$_{2.5}$ levels compared to reductions in other gas precursors. 
Moreover, \cite{de2009sensitivity} showed that a targeted 50\% reduction in ammonia emissions results in a decrease of total PM$_{2.5}$ levels by up to 2.4 $\mu g / m^3$ in the Lombardy region. Additionally, \cite{pozzer2017impact} demonstrated that the nonlinear behaviour of the sulphate-nitrate-ammonia system influences the efficacy of PM$_{2.5}$ control strategies. 

Our study differs from the above literature, as we use a statistical spatiotemporal model to perform a "what-if" scenario analysis assessing the reduction in PM$_{2.5}$ concentration due to certain ammonia emission reduction scenarios. In \cite{fasso2023extent}, the statistical relation between observed PM$_{2.5}$ and NH$_3$ concentrations is modelled using a generalised least square approach able to handle spatial and temporal correlation. This is
based on the five monitoring stations in Lombardy where observed concentrations of both substances are available. 

Unfortunately, the Lombardy air quality monitoring network is not specifically designed to monitor agricultural air pollutants. Most monitoring stations are located in areas with low NH$_{3}$ emission levels, and none are located where such emissions reach the highest peaks.
Consequently, only very few NH$_3$ concentration stations are available in the Lombardy region, and a fully spatiotemporal model is not viable on those data.
Moreover, such network spatial unbalance may imply a preferential sampling bias.

Over the last few decades, significant progress has been made in spatiotemporal statistical models in general \cite{cressie2010fixed, heaton2019, jurek2023, rue2009approximate}, and for air quality in particular \cite{calculli2015, Zheng2021, Zhang2023}.  
A well-established framework to study spatiotemporal processes is the state space model and the related Kalman filter technique \cite{calculli2015, fasso2010unified, huang1996spatio, Padilla2020, rougier2023}. In this context, \cite{calculli2015} proposes a multivariate spatiotemporal statistical model named the Hidden Dynamic Geostatistical Model (HDGM), which is a two-level hierarchical model suitable for complex environmental processes. In \cite{AgrimoniaComparions}, the HDGM has been compared with the Generalised Additive Mixed Model and the Random Forest with Residual Kriging, in PM$_{2.5}$ modelling in Lombardy. The HDGM demonstrated better performance in cross-validation. In addition, considering the heteroskedasticity, the authors point out that model uncertainty varies throughout the year for all three models, and PM$_{2.5}$ concentrations in winter can be predicted less accurately than in summer. 

In spatiotemporal models, heteroskedasticity has various facets and may refer to time, space, data heterogeneity or a mixture of the three. Also, the skedastic function may be deterministic or stochastic. 
An example of data heterogeneity arises in data fusion problems where the data vector elements are obtained by different sensors or processes. See, e.g., \cite{Smith2008}.

Considering the temporal dimension, we may model stochastic heteroskedasticity using the well-known approach introduced by the Nobel Prize Robert F. Engle \cite{Engle1982}. It is based on conditioning the error variance on the past and resulted in the large suite of GARCH-like models developed in the last decades. See, e.g., \cite{francq2019garch}. Also, the approach based on a deterministic skedastic function is used in environmental statistics, for example, a seasonal variance. See, e.g., \cite{Benth2007,fasso2023extent}.

Considering the spatial dimension, deterministic spatial skedastic functions have been used extensively. For example, it is considered by  \cite{MUR2009} in spatial econometrics models and by \cite{Hulshof2023} in ecology. Also, Engle and Bollerslev's generalised conditionally heteroskedastic approach has been recently introduced to spatial econometrics by \cite{OTTO2018}. For a review and developments, see \cite{Otto2023} and references therein.

In our study, we are not interested in modelling and interpreting the skedastic function per se but as a nuisance parameter needed to make correct inferences on spatial maps and aggregated results. For these reasons, we opt for a very flexible unstructured time-varying error variance applied to the HDGM. This results in a high-dimensional parameter dimension, which is efficiently handled by the expectation-maximisation (EM) algorithm. 

The rest of the paper is organised as follows. Section \ref{sec:methodology} discusses the methodology adopted. In particular Section \ref{sec:methodology_model} defines the heteroskedastic HDGM and the maximum likelihood estimation algorithm. Section \ref{sec:methodology_scenario_analysis} presents the "what-if" scenario analysis approach and provides the formula to compute the impact uncertainty; Section \ref{sec:CaseStudy} addresses the application of the heteroskedastic HDGM to the observed daily PM$_{2.5}$ in the Lombardy Region, between 2016 and 2020, and outlines the scenario analysis to assess the PM$_{2.5}$ changes due to NH$_3$ emissions; the results are summarised in Section \ref{sec:Discussion}. The Conclusions Section closes the paper.

\section{Methodology} \label{sec:methodology}
Despite the HDGM's capability to handle multivariate spatiotemporal data, this study focuses on the univariate case. In particular, the next section introduces the heteroskedastic HDGM to estimate PM$_{2.5}$ concentrations based on the data observed within the air quality monitoring network. 

\subsection{The Heteroskedastic Hidden Dynamic Geostatistical Model} \label{sec:methodology_model}

To understand the relationship between predictors and the response variable, taking into account spatial and temporal correlation, we propose a heteroskedastic extension of the univariate HDGM \cite{calculli2015, cressie2015} which is a two-level hierarchical model. The hierarchy is constructed by putting together two conditional submodels. At the first level, the observation variability is modelled by the measurement equation, which is essentially given by a regression component, a stochastic latent variable, and an error with time-varying variance. The latent variable is deﬁned at the second level of the hierarchy. It handles the spatiotemporal correlation through a Markovian process. The innovation term is a zero-mean Gaussian process with a spatial covariance function. 

Let $y(\bm{s},t)$ be the response variable observed at site $\bm{s} \in \mathbb{S}^2$, where $\mathbb{S}^2$ is the surface of the sphere embedded in $\mathbb{R}^3$, and discrete time $t = 1,...,T$. The heteroskedastic univariate HDGM is defined as follows:
\begin{equation}
    \begin{split}
           & y(\bm{s},t) = \bm{x}'(\bm{s},t)\bm{\beta} + \alpha z(\bm{s},t) + \epsilon(\bm{s},t) \\
           & z(\bm{s}, t) = g z(\bm{s}, t - 1) + \eta(\bm{s},t).
    \end{split}
\label{eq:hdgm_model}
\end{equation}

\noindent The $\bm{\beta}$ is a vector of fixed effect coefficients; $\bm{x}(\bm{s},t)$ is a $p \times 1$ vector of covariates that accounts for all exogenous effects; $\alpha$ is a scale parameter of the latent variable; the heteroskedastic measurement error $\epsilon(\bm{s},t) \sim N(0,\sigma^2_{\epsilon,t})$ is independent in space and time; $z(\bm{s},t)$ is a unit-variance Markovian scalar process ruled by the transition coefficient $g$; the innovation term $\eta(\bm{s},t)$ is a zero-mean Gaussian process, $GP(0, \rho(\|\bm{s} - \bm{s}'\|;\theta))$, independent in time where $\rho$ is a valid spatial correlation function and $\|\bm{s} - \bm{s}'\|$ is the geodetic distance between $\bm{s}$ and $\bm{s}' \in  \mathbb{S}^2$. The model parameters set is $\bm{\Psi} =\{ \bm{\beta}, g, \theta, \alpha, \sigma^2_{\epsilon,1}, ...,\sigma^2_{\epsilon,T} \} $ which is estimated using the maximum likelihood (ML) estimation through the EM algorithm. 

\subsection{Matrix representation of univariate heteroskedastic HDGM} \label{sec:methodology_matrix_model}
Using the same notation introduced in \cite{calculli2015}, suppose that the variable $y(\bm{s},t)$ is observed at each spatial location $\mathcal{S}_n = \{ \bm{s}_1,..., \bm{s}_n\}$. Let $\bm{y}_t = \big (y(\bm{s}_1,t),...,y(\bm{s}_n,t)\big)'$ be the $n \times 1$ vector of the response. With these assumptions, the process can be considered as a classical state-space model \citep{shumway2000time} where the observations at time $t$ follow the equations:
\begin{equation}
\label{eq:hdgm_matrix}
    \begin{split}
        & \bm{y}_t = \bm{X}_t \bm{\beta} + \alpha \bm{z}_t + \bm{\epsilon}_t \\
        & \bm{z}_t = g \bm{z}_{t-1} + \bm{\eta}_t 
    \end{split}
\end{equation}

\noindent with $\bm{X}_t = \big ( \bm{x}(\bm{s}_1,t),..., \bm{x}(\bm{s}_n,t)\big )'$. Vectors $\bm{z}_t$ and $\bm{\eta}_t$ are defined similarly to $\bm{y}_t$. Moreover, $\bm{\epsilon}_t$ is a $n$-dimensional random noise vector, $\bm{\epsilon}_t \sim N_n(\bm{0}, \bm{\Sigma}_{\epsilon,t})$ where $\bm{\Sigma}_{\epsilon,t}$ is given by $\bm{\Sigma}_{\epsilon,t} = \sigma_{\epsilon,t}^2 \bm{I}_n$, and $\bm{I}_n$ is the identity matrix of order $n$. The distribution of the latent variable at $t = 0$, $\bm{z}_0 \sim N_n(\bm{\mu}_0,\bm{\Sigma}_0)$. If all parameters in $\bm{\Psi}$ are known, the unobserved temporal process $\bm{z}_t$ in the model (\ref{eq:hdgm_matrix}) is estimated for each time $t$ through the Kalman smoother technique, initialised with the condition $\bm{z}_0$. Specifically, the  Kalman smoother yields the state estimate $\bm{z}_t^{T} = E_{\bm{\Psi}}(\bm{z}_t | \bm{Y})$ and the corresponding uncertainty $\bm{P}_t^{T} = \text{Var}_{\bm{\Psi}}(\bm{z}_t | \bm{Y})$. The Kalman smoother algorithm handles in a natural way time-varying parameters \cite{shumway2000time}. Alongside the Kalman smoother output, the quantities $\bm{S}_{10}$, $\bm{S}_{00}$, and $\bm{S}_{11}$ are introduced as the so-called EM second moments \cite{shumway2000time}.

\subsection{Complete-data likelihood}
From the model assumptions, we have the following probability distributions: 
\begin{equation*}
    \begin{split}
        & \bm{y_t} | \bm{z_t} \sim N_n(\bm{\mu}_t, \bm{\Sigma}_{\epsilon,t})  \\
        & \bm{z_t} | \bm{z}_{t-1} \sim N_n (g \bm{z}_{t-1}, \bm{\Sigma}_\eta) \\
    \end{split}
\end{equation*}

\noindent where $\bm{\mu}_t = \bm{X}_t \bm{\beta} + \alpha \bm{z}_t$. Following the results in Appendix A.1 of  \cite{calculli2015} and assuming that $\bm{\Sigma}_{\epsilon,t}$ is positive defined, the complete-data log-likelihood function for observations $\bm{Y} = \{\bm{y}_1, ... , \bm{y}_T\}$ and $\bm{Z} = \{\bm{z}_0, \bm{z}_1,... ,\bm{z}_T\}$ is given by
\begin{equation}
\label{eq:log_likelihood}
\begin{split}
    -2l(\bm{\Psi}; \bm{Y},\bm{Z}) & = \sum_{t = 1}^{T}{log|\bm{\Sigma}_{\epsilon,t}|} + \sum_{t = 1}^{T}{\bm{e}_t' \bm{\Sigma}_{\epsilon,t}^{-1} \bm{e}_t}  \\
    & + log|\bm{\Sigma}_0| + (\bm{z}_0 - \bm{\mu}_0)'\bm{\Sigma}_0^{-1}(\bm{z}_0 - \bm{\mu}_0) \\
    & + T log|\bm{\Sigma}_\eta| + \sum_{t = 1}^{T}(\bm{z}_t - g \bm{z}_{t-1})'\bm{\Sigma}_\eta^{-1}(\bm{z}_t - g \bm{z}_{t-1})
\end{split}
\end{equation}

\noindent where $\bm{e}_t = \bm{y}_t - \bm{\mu}_t$ and $|\cdot|$ is the matrix determinant. Due to the additive structure of Eq. (\ref{eq:log_likelihood}) where the right-hand terms depend on different subsets of the parameters, $l(\bm{\Psi}; \bm{Y}, \bm{Z})$ can be written as $l(\bm{\Psi}) = l(\bm{\Psi}_1) + l(\bm{\Psi}_0) + l(\bm{\Psi}_2)$ where $\bm{\Psi}_1= \{ \alpha, \bm{\beta},\sigma^2_{\epsilon,1}, ...,\sigma^2_{\epsilon,T}\}$, $\bm{\Psi}_0 = \{\bm{\mu}_0, \bm{\Sigma}_0 \} $ and $\bm{\Psi}_2 = \{ \theta, g\}$.

\subsection{Estimation formulas}
The ML estimation of the unknown parameter vector $\bm{\Psi}$ is performed using the iterates EM algorithm. At each iteration, there are two steps, the E-step and the M-step. The E-step ﬁnds the conditional expectation of the complete-data log-likelihood given the observation, namely 
\begin{equation*}
\label{eq:EM_q_function}
    Q(\bm{\Psi}, \bm{\Psi}^{(m)}) = E_{\bm{\Psi}^{(m)}}(-2l(\bm{\Psi}; \bm{Y}, \bm{Z}) | \bm{Y}) 
\end{equation*}

\noindent where $E_{\bm{\Psi}^{(m)}}$ is the expectation given the parameter estimate at iteration $m$. At the M-step, $Q(\bm{\Psi}, \bm{\Psi}^{(m)})$ is maximised with respect to $\bm{\Psi}$ and the new estimate is $\bm{\Psi}^{(m+1)} = \operatorname*{argmax}_{\bm{\Psi}} Q(\bm{\Psi}, \bm{\Psi}^{(m)})$. Due to the linear properties of the conditional expectation we can write $Q(\bm{\Psi}, \bm{\Psi}^{(m)}) = Q(\bm{\Psi}_1, \bm{\Psi}^{(m)})+ Q(\bm{\Psi}_0, \bm{\Psi}^{(m)}) + Q(\bm{\Psi}_2, \bm{\Psi}^{(m)})$. In this way, the maximisation step can be broken into several smaller optimisation problems. 

In the following, let $E(\cdot | \cdot) \equiv E_{\bm{\Psi}^{(m)}}(\cdot | \cdot)$ and $\text{Var}(\cdot | \cdot) \equiv \text{Var}_{\bm{\Psi}^{(m)}}(\cdot | \cdot)$. Moreover, $\bm{\mu}_0 \equiv  \bm{\mu}_0^{(m)}$, $\bm{\Sigma}_{0} \equiv \bm{\Sigma}_{0}^{(m)}$, $g \equiv g^{(m)}$, $\bm{\Sigma}_{\epsilon,t} \equiv \bm{\Sigma}_{\epsilon,t}^{(m)}$ and $\bm{\Sigma}_{\eta} \equiv \bm{\Sigma}_{\eta}^{(m)}$ that is, vectors and matrices are evaluated using the estimate parameters at iteration $m$ of the EM algorithm.

The $Q(\bm{\Psi}_i, \bm{\Psi}^{(m)})$ term for $i = 0, 1, 2$ is expressed as:
\begin{equation*}
\label{eq:conditional_expectation_1}
    \begin{split}
        & Q(\bm{\Psi}_1, \bm{\Psi}^{(m)}) = \sum_{t = 1}^{T} log|\bm{\Sigma}_{\epsilon,t}| \\
        & + tr\Big(\sum_{t = 1}^{T}E(\bm{e}_t|\bm{Y})'\bm{\Sigma}_{\epsilon,t}^{-1}E(\bm{e}_t|\bm{Y}) + \bm{\Sigma}_{\epsilon,t}^{-1} \text{Var}(\bm{e}_t | \bm{Y}) \Big)
    \end{split}
\end{equation*}


\begin{equation*}
    \begin{split}
        & Q(\bm{\Psi}_0, \bm{\Psi}^{(m)}) =  log|\bm{\Sigma}_{0}| \\
        & + tr\Big[ \bm{\Sigma}_{0}^{-1}  \Big( E(\bm{z}_0|\bm{Y}) - \bm{\mu}_0\Big) \Big( E(\bm{z}_0|\bm{Y}) - \bm{\mu}_0\Big)' + \text{Var}(\bm{z}_0 | \bm{Y}) \Big]
    \end{split}
\end{equation*}

\begin{equation*}
    \begin{split}
        Q(\bm{\Psi}_2, \bm{\Psi}^{(m)}) =  T log|\bm{\Sigma}_{\eta}|  + tr\Big[ \bm{\Sigma}_{\eta}^{-1}  \Big( \bm{S}_{11} - 2g\bm{S}_{10} + g^2\bm{S}_{00}  \Big) \Big]
    \end{split}
\end{equation*}

\noindent where $E(\bm{e}_t|\bm{Y})$ and $\text{Var}(\bm{e}_t|\bm{Y})$ are given in Appendix A.1 and Appendix A.2 of \cite{calculli2015} respectively. The maximisation step for updating $\bm{\Psi}_i$ is derived by solving 
\begin{equation*}
\begin{split}
    \frac{\partial}{\partial \bm{\Psi}_i} Q_i(\bm{\Psi}_i, \bm{\Psi}^{(m)}) = 0.
\end{split}
\end{equation*}

As a result, starting with initial value $\bm{\Psi}^{(0)}$, the updating formulas are: 
\begin{equation}
\label{eq:updating_epselon_1}
    (\sigma_{\epsilon,t}^2)^{(m+1)} = \frac{1}{n} tr(\bm{\Omega}_t^{(m)})
\end{equation}

\begin{equation}
\label{eq:updating_beta_2}
\begin{split}
    & \bm{\beta}^{(m+1)} = \Big [\sum_{t = 1}^{T} (\bm{X}_t)' (\bm{\Sigma}_{\epsilon,t}^{(m)})^{-1} \bm{X}_t \Big ]^{-1} \\
    & \Big [\sum_{t = 1}^{T} (\bm{X}_t)' (\bm{\Sigma}_{\epsilon,t}^{(m)})^{-1} (\bm{y}_t - \alpha^{(m)} \bm{z}_t^{T,(m)}) \Big ] 
\end{split}
\end{equation}

\begin{equation}
\label{eq:updating_alpha_3}
\begin{split}
    & \alpha^{(m+1)} = \sum_{t = 1}^{T} tr \Big[ \bm{z}_t^{T,(m)} (\bm{\Sigma}_{\epsilon,t}^{(m)})^{-1} \Big (\bm{y}_t - \bm{X}_t \bm{\beta}^{(m)} \Big ) \Big ] \\
    & \Big( \sum_{t = 1}^{T} tr \Big[ (\bm{\Sigma}_{\epsilon,t}^{(m)})^{-1} \Big (\bm{z}_t^{T,(m)} (\bm{z}_t^{T,(m)})' + \bm{P}_t^{T,(m)} \Big )\Big ] \Big)^{-1}
\end{split}
\end{equation}

\begin{equation}
\label{eq:updating_mu_4}
    \bm{\mu}_0^{(m+1)} = \bm{z}_0^{T,(m)}
\end{equation}

\begin{equation}
\label{eq:updating_sigma_5}
    \bm{\Sigma}_0^{(m+1)} = \bm{P}_0^{T,(m)}
\end{equation}

\begin{equation}
\label{eq:updating_g_6}
    g^{(m+1)} = tr(\bm{S}_{10}^{(m)}) tr(\bm{S}_{00}^{(m)})^{-1}
\end{equation}

\begin{equation}
\label{eq:updating_theta_7}
\begin{split}
    & \theta^{(m+1)} = \operatorname*{argmax}_{\theta} \Big(  T log|\bm{\Sigma}_{\eta}^{(m)}| \\
    & + tr\Big[ (\bm{\Sigma}_{\eta}^{(m)})^{-1}  \Big( \bm{S}_{11}^{(m)} - 2g^{(m)}\bm{S}_{10}^{(m)} + (g^{(m)})^2\bm{S}_{00}^{(m)}  \Big) \Big] \Big) 
\end{split}
\end{equation}

\noindent where $\bm{\Omega}_t^{(m)} = E(\bm{e}_t|\bm{Y})E(\bm{e}_t|\bm{Y})' + \text{Var}(\bm{e}_t | \bm{Y})$ is given in Appendix A.3 of \cite{calculli2015}. Note that the updating formulas (\ref{eq:updating_epselon_1}) to (\ref{eq:updating_theta_7}), are similar to those provided by \cite{calculli2015}. The main exception lies in the parameters in $\bm{\Psi}_0$ which defines the measurement equation and undergoes modification due to the introduction of time-varying variance. Specifically, Eq. (\ref{eq:updating_epselon_1}) results in an average over space for each time $t$. The updating formulas for $\bm{\beta}$ and $\alpha$, in Eqs. (\ref{eq:updating_beta_2}) and (\ref{eq:updating_alpha_3}) respectively, take into account the time-varying variance $\bm{\Sigma}_{\epsilon,t}$. 

Given the ML estimate $\bm{\hat{\Psi}}$, predictions of the response variable at new sites $\bm{s}_0$ and time $t = 1,..., T$ are given by
\begin{equation}
\label{eq:hdgm_kriging}
    \hat{y}(\bm{s}_0,t) = \bm{x}'(\bm{s}_0,t)\bm{\hat{\beta}} + \hat{\alpha} z_t^T(\bm{s}_0). \\
\end{equation}

\subsection{Scenario analysis approach} \label{sec:methodology_scenario_analysis}

This section presents the scenario analysis approach adopted to assess the PM$_{2.5}$ sensitivity to the NH$_3$ emissions. We consider the "What-if" scenario analysis approach: "What would happen if the NH$_3$ reductions were fully implemented at time 0 (1st January 2016)?".  We compare the predicted PM$_{2.5}$, $\hat{y}(\bm{s},t)$, based on the observed NH$_3$ emissions with a prediction, $\hat{y}^r(\bm{s},t)$ where the NH$_3$ emissions are reduced by a factor $r$. The daily PM$_{2.5}$ change $\Delta_{y}(\bm{s},t) = y(\bm{s},t) - y^r(\bm{s},t)$ and its variance are estimated by 
\begin{equation*}
\begin{split}
    & \Delta_{\hat{y}}(\bm{s},t) = \hat{y}(\bm{s},t) - \hat{y}^r(\bm{s},t) \\
    & \text{Var}(\Delta_{\hat{y}}(\bm{s},t) - \Delta_{y}(\bm{s},t)) = \Delta_{\bm{x}}' \bm{\Sigma}_{\hat{\bm{\beta}}} \Delta_{\bm{x}} + \text{Var}(\epsilon_{t} - \epsilon_{t}^r)
\end{split}
\label{eq:daily_reduction}
\end{equation*}

\noindent where $\Delta_{\bm{x}} = \bm{x}(\bm{s},t) - \bm{x}^{r}(\bm{s},t)$; $\bm{\Sigma}_{\hat{\bm{\beta}}}$ is the estimate of the $\bm{\hat{\beta}}$ covariance matrix;  $\epsilon_{t}$ and $\epsilon_{t}^r$ are the measurement errors of both predictions which are independent and with the same variance. The average of the daily $\Delta_{\hat{y}}(\bm{s},t)$ and its variance over space subset $\mathcal{D}$ and time interval $\mathcal{I}$ are given by  
\begin{equation}
        \bar{\Delta}_{\hat{y}} = \frac{1}{\mathcal{D}^{*} \mathcal{I}^{*}} \sum_{\bm{s} \in \mathcal{D}} \sum_{t \in \mathcal{I}} \Delta_{\hat{y}}(\bm{s},t) \\
\label{eq:aggregation_delta_mean}
\end{equation}
\begin{equation}
    \text{Var}(\bar{\Delta}_{\hat{y}} - \bar{\Delta}_{y}) = \bar{\Delta}_{\bm{x}}' \bm{\Sigma}_{\hat{\bm{\beta}}} \bar{\Delta}_{\bm{x}} + \frac{2}{(\mathcal{I}^{*})^2 \mathcal{D}^{*}} \sum_{t \in \mathcal{I}} \hat{\sigma}_{\epsilon,t}^2  
\label{eq:aggregation_delta_var}
\end{equation}

\noindent where $\mathcal{D}^{*}$ and $\mathcal{I}^{*}$ are the number of pixels and days involved in the average computation;  $\hat{\sigma}_{\epsilon,t}^2$ is the estimate of the variance of the measurement error and $\bar{\Delta}_{\bm{x}}$ is the average of the design matrix. The Eq. (\ref{eq:aggregation_delta_var}) provides the standard deviation ($\text{std}$) used to assess the estimate change uncertainty in the next session.

\section{Lombardy case study} \label{sec:CaseStudy}
In this section, the methodology discussed in Section \ref{sec:methodology} is used to estimate the impact of NH$_3$ emissions on the concentration of PM$_{2.5}$ in the Lombardy Plain. The dataset description and preliminary analysis are detailed in Section \ref{sec:Data_Description} and Section \ref{sec:preliminary_analysis}, while the model estimation is described in Section \ref{sec:model_result}. Finally, Section \ref{sec:Scenario_Analysis_case_study} discuss the scenario analysis implementation, details the result and provides maps of the PM$_{2.5}$ change.

\subsection{Data description} \label{sec:Data_Description}
Our research is based on the Agrimonia dataset \cite{AgrimoniaDataset}, which is a comprehensive daily spatiotemporal dataset for modelling the air quality of the Lombardy region. This dataset is open access and is available through the Zenodo repository \cite{AgrimoniaZenodo}. It covers the period from 2016 to 2021 and includes daily concentrations of air pollutants, meteorological conditions, emission fluxes, land use, and livestock densities. Because spatiotemporal methods can benefit from neighbouring information to improve prediction performance near boundaries, the dataset also provides data for an area around the Lombardy region, obtained by applying a 0.3$^\circ$ buffer around regional boundaries. 

We consider the period between 2016 and 2020 and the following variables: PM$_{2. 5}$ measured at 45 ground stations (30 of which belong to the Lombardy region, while the remaining 15 are located in the neighbouring area), wind speed (average wind speed at 100 $m$), temperature (air temperature at 2 $m$), relative humidity (RH), boundary layer height (BLH, maximum daily air mixing layer height), high vegetation index (HVI, high vegetation abundance), sulphur dioxide emissions (SO$_2$), total ammonia emissions from agriculture (NH$_3$), and nitrogen oxide emissions (NO$_x$). 

In addition, we create new binary variables: Rain $= 1$ if the total daily precipitation exceeds the threshold of 1 $mm$, Urban $=1$ if land use is classified as urban, and a new categorical variable Season which takes four categories: \textit{Winter}, \textit{Spring}, \textit{Summer}, and \textit{Autumn}. Table \ref{tab:covariates} summarises the variables used in this study along with the main statistics. 

\begin{table}[h]
\caption{Variables selected from the Agrimonia dataset \cite{AgrimoniaDataset} and main descriptive statistics. The Rain and Urban variables are binary. Season is not included because it is a categorical variable.}
\label{tab:covariates}
\begin{tabular}{lrrrr}
\toprule
Name & min & mean &  max &std  \\ 
\midrule
PM$_{2.5}$ $[\mu g / m^3]$ & 1.00 & 21.58 &  182.00 &16.63 \\
Wind speed $[m/s]$ & 0.56  & 2.59 &  11.93 &1.33 \\
Temperature $[C^{\circ}]$  & -11.94  & 13.44 &  32.88 &7.98  \\
RH  $[\%]$          & 25.87   & 74.34 &  99.10 &12.26 \\
BLH $[m]$          & 27.38   & 1043.91 &  4420 &556.77  \\
HVI $[m^2/m^2]$          & 0.86  & 2.23 &  5.03 &0.70 \\
SO$_2$ $[mg/(m^2 day)]$         & 0.05 & 3.34 &  45.06 &5.53 \\
NH$_3$ $[mg/(m^2 day)]$        & 0.13  & 11.55 &  72.03 &12.00  \\
NO$_x$ $[mg/(m^2 day)]$        & 0.75  & 14.06 &  118.60 &16.73 \\
Rain & 0 & 0.02 & 1 & 0.15\\
Urban & 0 & 0.75 & 1 & 0.43 \\
\bottomrule
\end{tabular}
\end{table}

\subsection{Preliminary analysis} \label{sec:preliminary_analysis}
It is well-noted that the air quality data are characterised by a seasonal pattern \citep{lonati2020temporal, AgrimoniaComparions}. In particular, the PM$_{2.5}$ concentrations are higher in winter, due to meteorological conditions that reduce air circulation. The NO$_x$ emissions follow human activities and during the cold months reach high levels of emissions due to endothermic-powered vehicles and domestic heating systems. Vice versa the seasonal behaviour of the NH$_3$ emissions is determined by seasonal practices of agricultural activities responsible for strong variations in NH$_3$ emissions level. In particular, the NH$_3$ emissions are higher in the Spring and Summer periods. Table \ref{tab:seasonality} summarises the daily temperature average, emissions flow, and PM$_{2.5}$ by season showing the composition regimes in the atmosphere.

\begin{table}
    \centering
    \caption{Seasonal average of daily Temperature $[C^{\circ}]$, NH$_3$, NO$_x$ and SO$_2$ emissions $[mg/m^2]$ and PM$_{2.5}$ concentrations $[\mu g / m^3]$.}
    \label{tab:seasonality}
    \begin{tabular}{lrrrrr}
    \toprule
    Season & Temperature & NH$_3$ & NO$_x$ & SO$_2$ & PM$_{2.5}$ \\
    \midrule
    Winter & 3.88 & 5.21 & 19.97 & 4.11 & 35.87 \\
    Spring & 12.67 & 17.45 & 11.33 & 2.98 & 16.54 \\
    Summer & 23.23 & 14.71 & 10.45 & 2.80 & 12.19 \\
    Autumn & 13.83 & 8.70 & 14.59 & 3.46 & 21.62 \\
    \bottomrule
    \end{tabular}
\end{table}

\begin{figure}[ht]
\centering
\includegraphics[scale=.30]{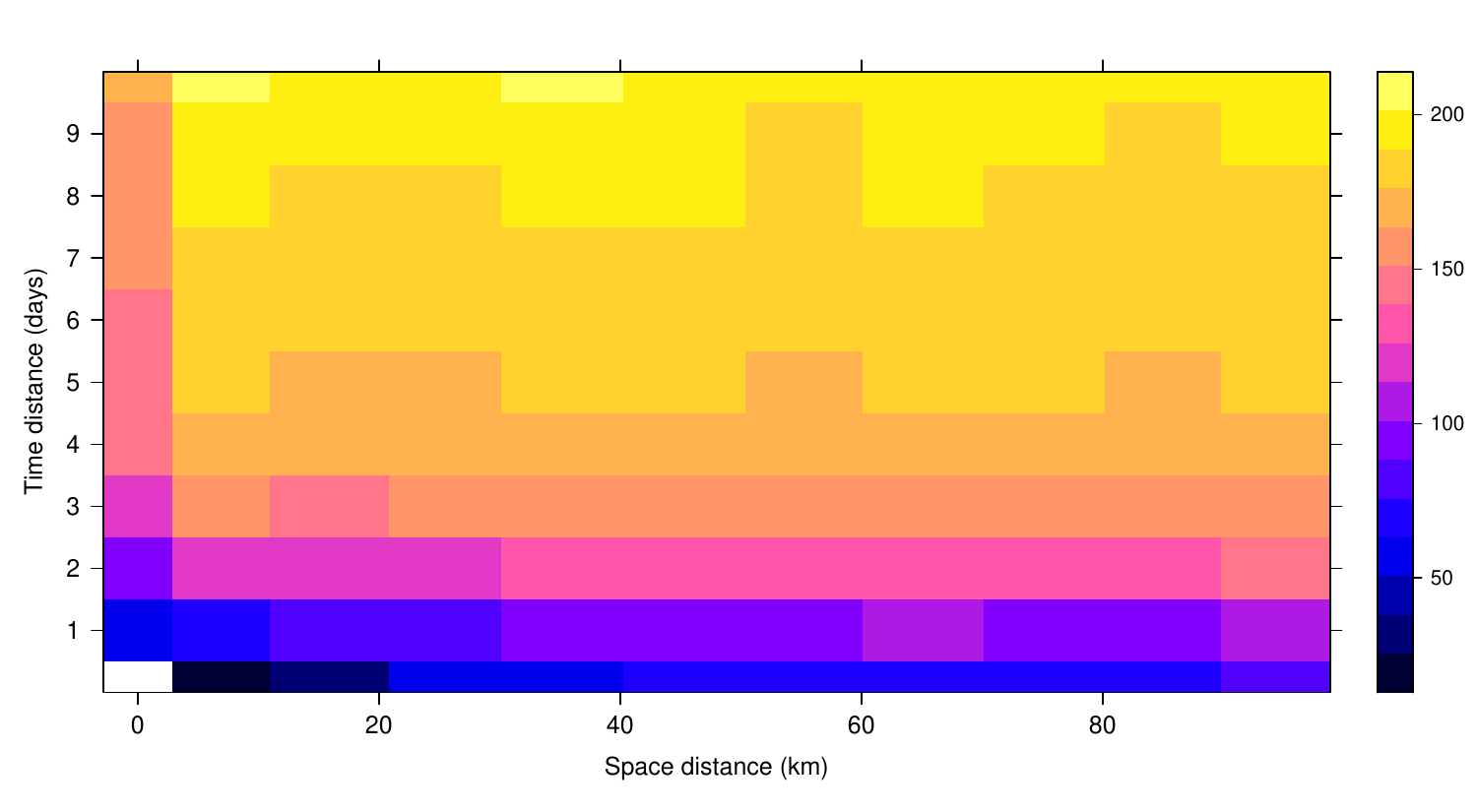}
\caption{Spatiotemporal variogram computed on PM$_{2.5}$ daily observations at 45 stations from 2016 to 2020.}
\label{fig:variogram}       
\end{figure}

The spatiotemporal variogram \cite{cressie2015} in Figure \ref{fig:variogram} highlights strong spatial and temporal correlation, as it shows that the variance increases with distance in both space and time.  The temporal dynamics was further analysed through the partial autocorrelation function, which highlights the first lag (1 day) as the most significant autocorrelated component, indicating that a model with first-order Markovian dynamics is appropriate. The autocorrelation at lag one is close to $0.8$ for all stations and suggests that PM$_{2.5}$ is relatively stable in the atmosphere. 

\begin{figure}[ht]
\centering
     \includegraphics[scale=0.65]{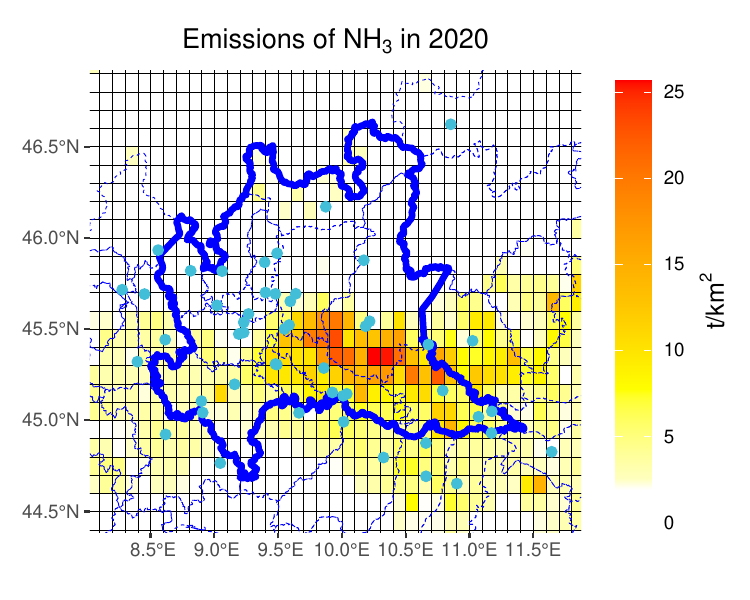}
    \caption{Annual NH$_3$ emissions ($t/km^2$) over the augmented Lombardy region in 2020. The cyan circles represent air quality stations. The continuous blue line represents the boundary of the Lombardy region while the dotted ones represent the province boundaries. \label{fig:emission_all}}
\end{figure}

A special focus is devoted to NH$_3$ emissions. The daily emission flux included in the Agrimonia dataset was provided monthly by the Copernicus Atmosphere Monitoring Service \cite{cams}. The spatial distribution of annual NH$_3$ emissions in 2020 is shown in Figure \ref{fig:emission_all}. The annual NH$_3$ emissions in 2020 aggregated by province are summarised in Table \ref{tab:nh3_provinces}. In particular, the provinces of Brescia, Cremona, and Mantua are characterised by higher emissions. 

It is worth noting that the air quality monitoring network is not specifically designed to monitor ammonia emissions. Indeed, most monitoring stations are located in areas with low NH$_{3}$ emission levels and none are located where they reach the highest peaks, as shown in Figure \ref{fig:emission_all}. Raising potential issues of preferential spatial sampling bias which is briefly discussed later in Section \ref{sec:Conclusions}.

\begin{table}[h]
    \centering
    \caption{Total NH$_3$ emissions by province in 2020.}
    \label{tab:nh3_provinces}  
\begin{tabular}{lcr}
\toprule
    Province     & Province code    & NH$_3$ $[t/year]$ \\
\midrule
Bergamo     & BG & 7546 \\
Brescia     & BS & 29524 \\
Como        & CO & 1517 \\
Cremona     & CR & 16892 \\
Lecco       & LC & 615 \\
Lodi        & LO & 6552 \\
Mantua      & MN & 20336 \\
Milan       & MI & 14013\\
Monza and Brianza & MB & 714 \\
Pavia       & PV & 8199 \\
Sondrio     & SO & 4601 \\
Varese      & VA & 1052 \\
\bottomrule
\end{tabular}
\end{table}

\subsection{Estimated model}
\label{sec:model_result}
Given the extensive literature on air quality in the Lombardy region, we rely on these studies to select the covariates for the model (\ref{eq:hdgm_model}). Specifically, \cite{AgrimoniaComparions} provides detailed insights into the impact of weather conditions on PM$_{2.5}$ formation in Lombardy, while \cite{fasso2023extent} explores the seasonality of pollutant concentration, revealing that the sensitivity of PM$_{2.5}$ to NH$_3$ and NO$_x$ varies with Season. Based upon these findings, we adopt a univariate HDGM with a regression term specified as follows 
\begin{multline*}
 PM_{2.5} \sim (Intercept) + Wind speed + Temperature \\
 + RH + Rain + BLH + Urban + HVI + SO_2  \\ 
 + NO_x + NH_3 + (NO_x + NH_3):Season    
\end{multline*}

\noindent while the spatial correlation function of the innovation term for the model (\ref{eq:hdgm_model}) is defined by the exponential function as $\rho(\|\bm{s} - \bm{s}'\|;\theta)) =  exp \{ - \|\bm{s} - \bm{s}'\| / \theta \}$.

\begin{figure}[ht]
    \centering
    \includegraphics[scale=0.6]{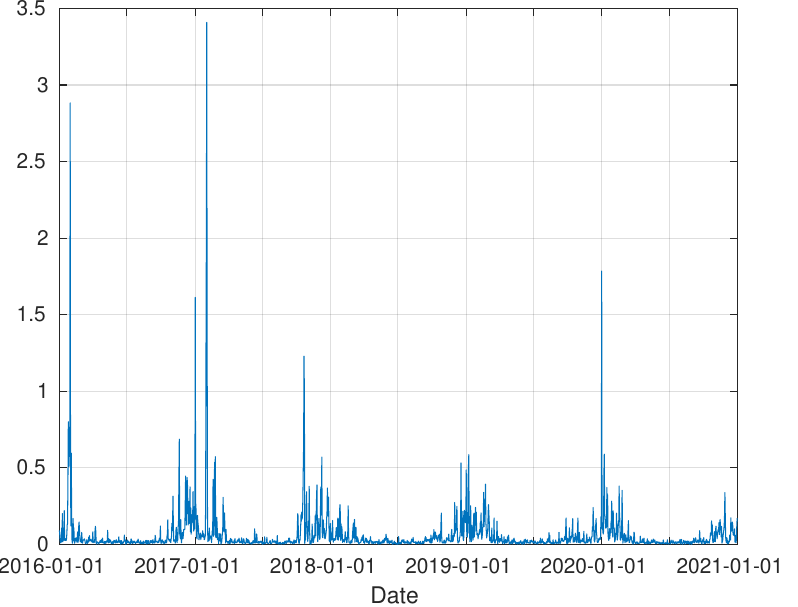}
    \caption{Estimated error variance, $\hat{\sigma}_{\epsilon,t}^2$. }
    \label{fig:sigma_eps_time}
\end{figure}

To avoid numerical issues, the response variable and covariates are standardised. So, the estimated parameters in the current subsection refer to this standardised setup. Table \ref{tab:HDGM_coefficient} summarises the estimated coefficients $\bm{\beta}$ of the linear regression model used for the large-scale component. All coefficients except the intercept are statistically significant. As expected, the role of NH$_3$ is most prominent in winter, when NH$_3$ plays a limiting role in the formation of PM$_{2.5}$. The estimated parameters of the latent variable and the uncertainty associated with them are $\hat{\theta} = 2.26 ^ \circ$ $(std < 0.06 ^ \circ)$, $\hat{g} = 0.79$ $(std <0.01)$ and $\hat{\alpha} = 0.18$ $(std <0. 01)$. Note that the latent process is stationary and the magnitude of $g$ indicates that the $z(\bm{s},t)$ change smoothly over time. Finally, Figure \ref{fig:sigma_eps_time} shows the daily $\hat{\sigma}_{\epsilon,t}^2$ which is higher in winter than in summer, and highlights the need for a heteroskedastic model. 

We validate the model by considering its performance in Cross-Validation (CV). In particular, we use the Leave-One-Station-Out CV scheme (LOSOCV), a variation of the commonly used leave-one-out CV approach applied in the spatiotemporal framework \cite{nowak2020}. To do this, we only use the 30 stations belonging to Lombardy in the LOSOCV procedure while the remaining 15 stations are used in the training set only. As a result, the model in-sample Root Mean Square Error (RMSE) is $3.85 \mu g/m^3$ while the CV-RMSE is $5.91 \mu g/m^3$, which improves the CTM performance \citep{veratti2023impact}.

\begin{table}[h]
\centering
\caption{Estimate of the fixed effect coefficients of the model (\ref{eq:hdgm_model}). $|\text{t}|$ is the absolute value of the $t$-statistic. }
\label{tab:HDGM_coefficient}
\begin{tabular}{lrrrr}
\toprule
Name & $\beta$ & std & |t| & p-value \\ 
\midrule
(Intercept)  & -0.02 & 0.05 & 0.46  & 0.64                \\
Wind speed    & -0.08 & 0.00 & 23.87 & 0                   \\
Temperature          & -0.21 & 0.02 & 10.63 & 0                   \\
RH             & 0.09  & 0.01 & 17.39 & 0                   \\
Rain           & -0.01 & 0.00 & 2.87  & 4.10E-03            \\
BLH            & -0.07 & 0.00 & 14.87 & 0                   \\
Urban          & -0.08 & 0.01 & 15.24 & 0                   \\
HVI            & -0.03 & 0.00 & 6.88  & 0            \\
SO2           & -0.05 & 0.01 & 9.13  & 0            \\
NO$_x$         & 0.10  & 0.01 & 10.95 & 0                   \\
NO$_x$:Winter  & 0.03  & 0.01 & 3.35  & 7.96E-04            \\
NO$_x$:Summer  & -0.02 & 0.01 & 3.88  & 1.06E-04              \\
NO$_x$:Spring  & -0.02 & 0.01 & 4.06  & 4.96E-05            \\
NH$_3$       & 0.11  & 0.01 & 8.96  & 0                   \\
NH$_3$:Winter  & 0.09  & 0.01 & 12.07 & 0                   \\
NH$_3$:Summer & -0.07 & 0.01 & 6.34  & 0            \\
NH$_3$:Spring & -0.10 & 0.01 & 7.87  & 0            \\
\bottomrule
\end{tabular}
\end{table}

\begin{figure}
    \centering
    \includegraphics[scale=0.6]{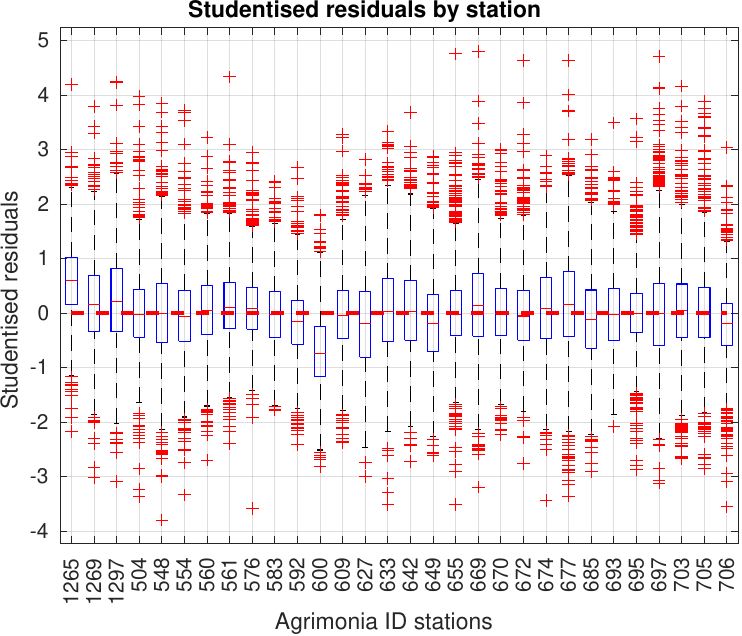}
    \caption{Studentised residuals grouped by station. The outliers (red cross) are identified by the Whisker length based on the first and third quartiles.}
    \label{fig:residual_boxplot}
\end{figure}

\begin{figure}
    \centering
    \includegraphics[scale=0.6]{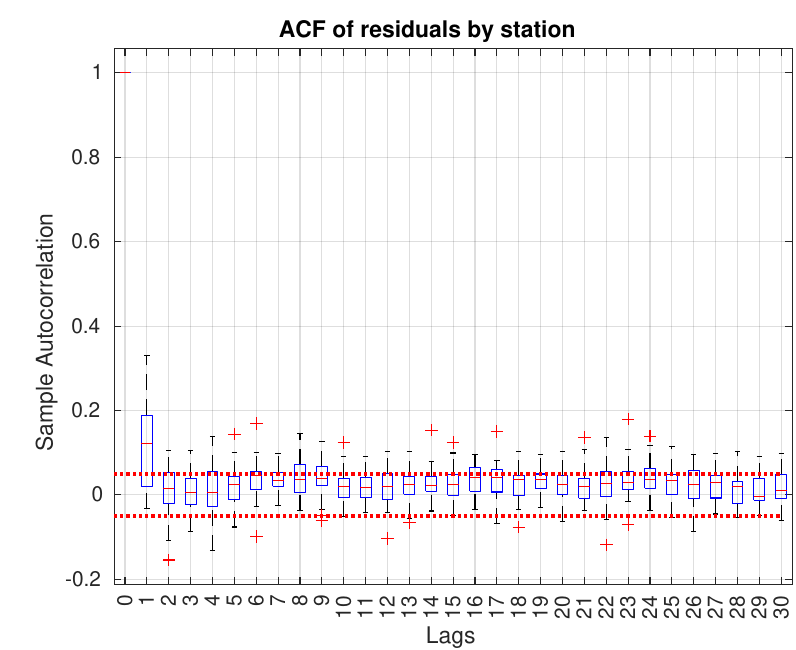}
    \caption{Autocorrelation Function (ACF) computed by station residuals for the first 30 lags (days) and summarised through boxplots. The outliers (red cross) are identified by the Whisker length based on the first and third quartiles.}
    \label{fig:acf_residual}
\end{figure}

In Figure \ref{fig:residual_boxplot}, the distributions of the studentised residuals, computed as $\bm{e}_t / \hat{\sigma}_{\epsilon,t}$, grouped by station are shown through boxplots. Their distributions are approximately centred around zero and moderately non-Gaussian. The residuals are further investigated through the Autocorrelation Function (ACF). We compute the ACF by station and then we summarise the autocorrelation coefficients for all stations by using the boxplot representation, as shown in Figure \ref{fig:acf_residual}. The ACF is generally not significant except at lag one where there is a weak autocorrelation. This is not further considered as its influence in Eq. (\ref{eq:aggregation_delta_var}) is negligible.   

\subsection{Scenario Analysis} \label{sec:Scenario_Analysis_case_study}

According to \cite{veratti2023impact} we propose two different scenarios for NH$_3$ emissions reduction: \textit{(i)} scenario named PRIA that accounts for a reduction of 26\% in NH$_3$ emission; \textit{(ii)} scenario named Strong that is characterised by a reduction of 50\% in ammonia emissions. The PRIA scenario is based on the PRIA plan (in Italian "Piano Regionale degli Interventi per la qualità dell'Aria"), which identifies the actions needed to reduce ammonia emissions by 26\% in Lombardy.  

We consider the "What-if" scenario analysis approach: "What would happen if the NH$_3$ reductions were fully implemented at time 0 (1st January 2016)?". Considering the distribution of NH$_3$ emissions shown in Figure \ref{fig:emission_all}, we assess the scenario analysis only in non-forested areas below 640 $m$ altitude. The results are back-transformed to the original units.

For each scenario, thanks to Eq. (\ref{eq:hdgm_kriging}), we map the average and the uncertainty of the PM$_{2.5}$ changes due to NH$_3$ reductions on a regular grid of 0.1$^\circ$ $\times$ 0.1$^\circ$ over the studied area. We focus on winter (452 daily predictions for each pixel), which is characterised by high PM$_{2.5}$ concentrations (see Table \ref{tab:seasonality}) and the biggest effect of NH$_3$ (see Table \ref{tab:HDGM_coefficient}). 

\begin{figure}[ht]
    \centering
    \subfloat[][]{
        \label{fig:Scenario_A_map}
        \includegraphics[scale=0.57]{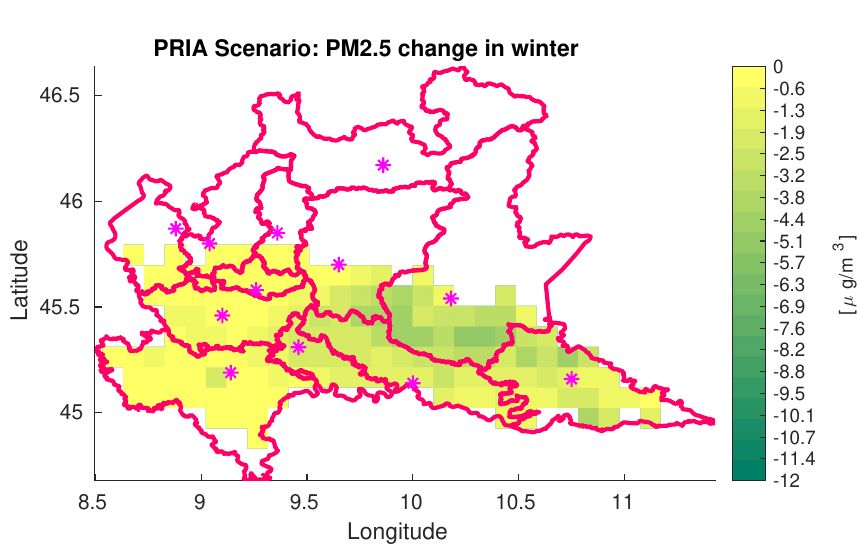}
        }%
    \qquad
    \subfloat[][]{   
        \label{fig:Scenario_A_map_std}
        \includegraphics[scale=0.57]{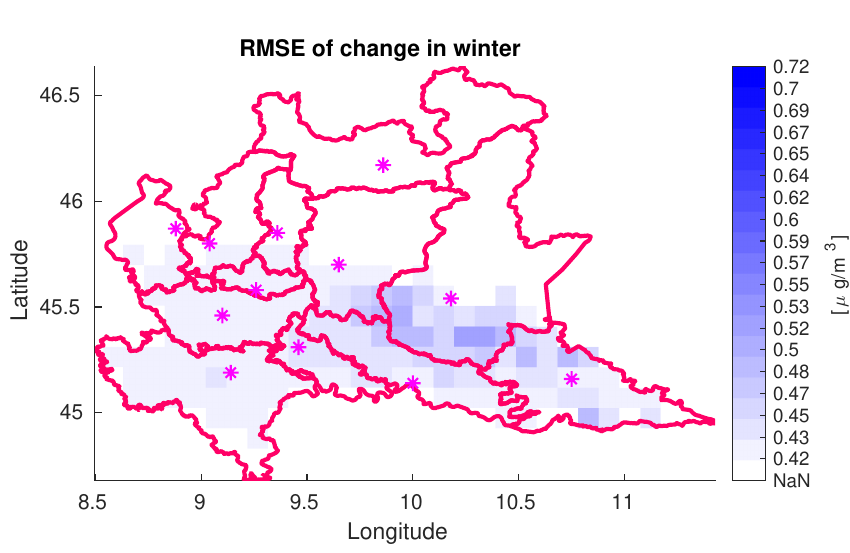}
        }%
    \caption{PRIA Scenario (-26\%). Panel a: average PM$_{2.5}$ reduction in winter, $\bar{\Delta}_{\hat{y}}$  (452 daily observations for each pixel). Panel b: uncertainty associated with the average reduction ($\text{std}(\bar{\Delta}_{\hat{y}})$). The pink stars depict the provincial capitals listed in Table \ref{tab:nh3_provinces}. The model is only evaluated in non-forested areas below 640 $m$ altitude.}%
    \label{fig:Scenario_A}%
\end{figure}

\begin{figure}[ht]
    \centering
    \subfloat[][]{
        \label{fig:Scenario_B_map}
        \includegraphics[scale=0.57]{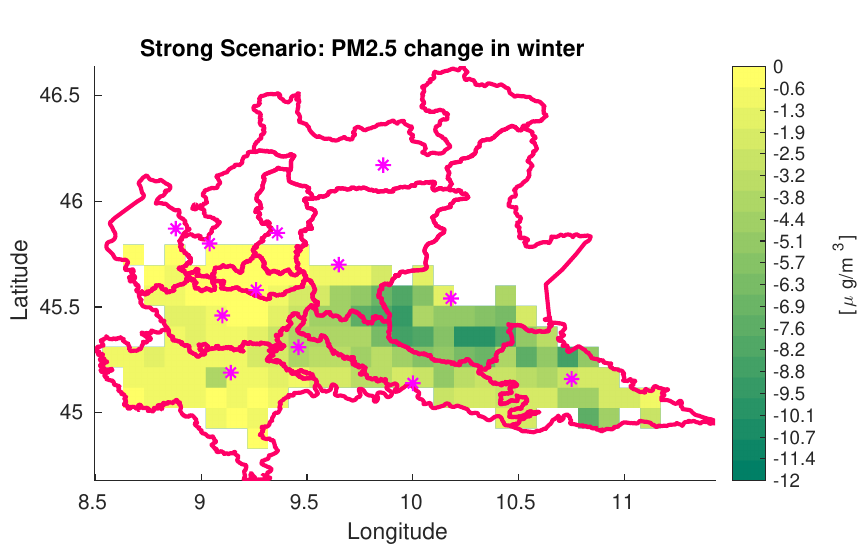}
        }%
    \qquad 
    \subfloat[][]{   
        \label{fig:Scenario_B_map_std}
        \includegraphics[scale=0.57]{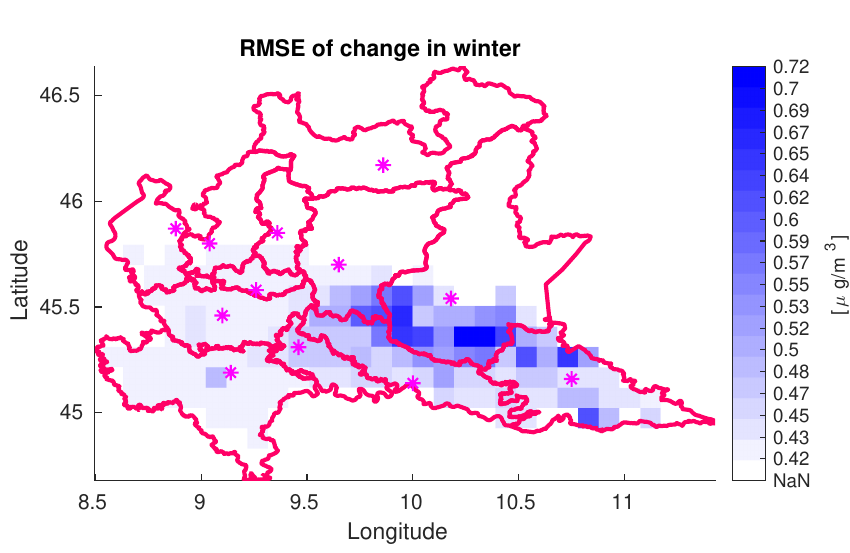}   
        }%
    \caption{Strong Scenario (-50\%). Panel a:  winter $\bar{\Delta}_{\hat{y}}$  (452 daily observations for each pixel). Panel b: uncertainty associated with the average reduction ($\text{std}(\bar{\Delta}_{\hat{y}})$). The pink stars depict the provincial capitals listed in Table \ref{tab:nh3_provinces}. The model is only evaluated in non-forested areas below 640 $m$ altitude.}%
    \label{fig:Scenario_B}%
\end{figure}

Considering winter, Figure \ref{fig:Scenario_A} depicts the average PM$_{2.5}$ reduction for the PRIA scenario, and the associated uncertainty. Analogously, Figure \ref{fig:Scenario_B} shows the average impact of the Strong scenario. Both scenarios show that in winter, the main reduction effect is obtained in the southeast area of the region where the NH$_3$ emissions are the highest. In particular, in some areas, the average reduction of PM$_{2.5}$ concentrations is close to 6 $\mu g /m^3$. Considering the province plain average, Brescia and Cremona, located in the southeast part of the region, show the highest PM$_{2.5}$ reductions of 2-3 $\mu g /m^3$. On the other hand, highly urbanised lands such as the metropolitan area of Milan, characterised by low levels of NH$_3$ emissions, do not benefit from the reduction of ammonia emissions. The more marked reduction of NH$_3$ emissions (Strong scenario) leads to more significant improvements in air quality, achieving local PM$_{2.5}$ reductions average close to 12 $\mu g /m^3$, reducing the risks associated with the health of the population. 

\begin{figure}[ht]
    \centering
    \subfloat[][]{
        \label{fig:Scenario_A_province_boxplot}
        \includegraphics[scale=0.57]{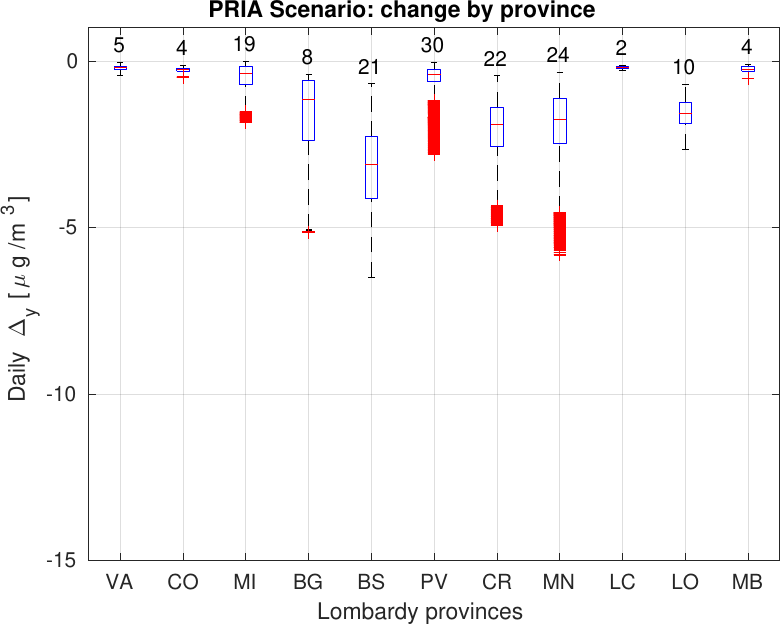}
        }%
    \qquad
    \subfloat[][]{   
        \label{fig:Scenario_A_land_boxplot}
        \includegraphics[scale=0.57]{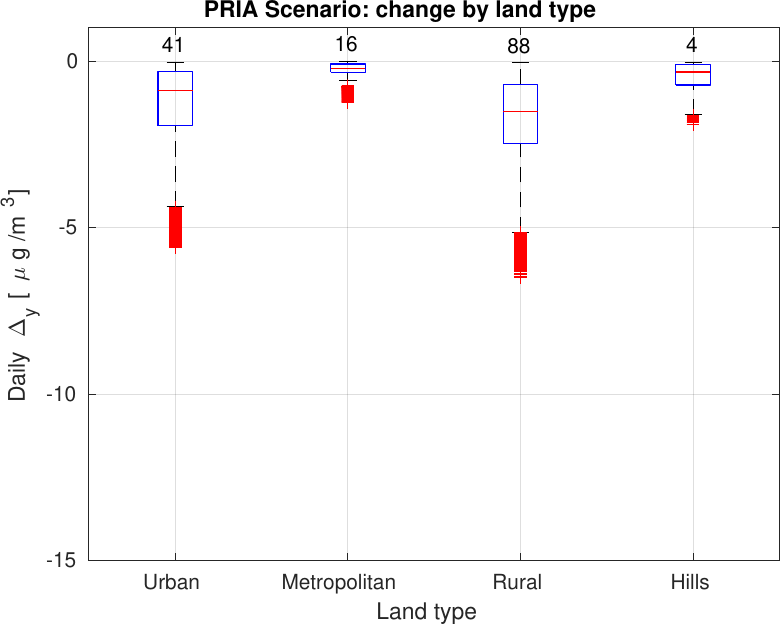}
        }%
    \caption{Scenario PRIA (-26\%). Boxplot of daily $\Delta_{\hat{y}}$ in winter (452 daily observations for each pixel). Panel a: grouped by province; panel b: grouped by land type. The number of pixels of each category is reported over the corresponding boxplot. The model is evaluated only in the non-forested area under 640 $m$ altitude.}
    \label{fig:Scenario_A_boxplot}
\end{figure}

\begin{figure}[ht]
    \centering
    \subfloat[][]{
        \label{fig:Scenario_B_province_boxplot}
        \includegraphics[scale=0.57]{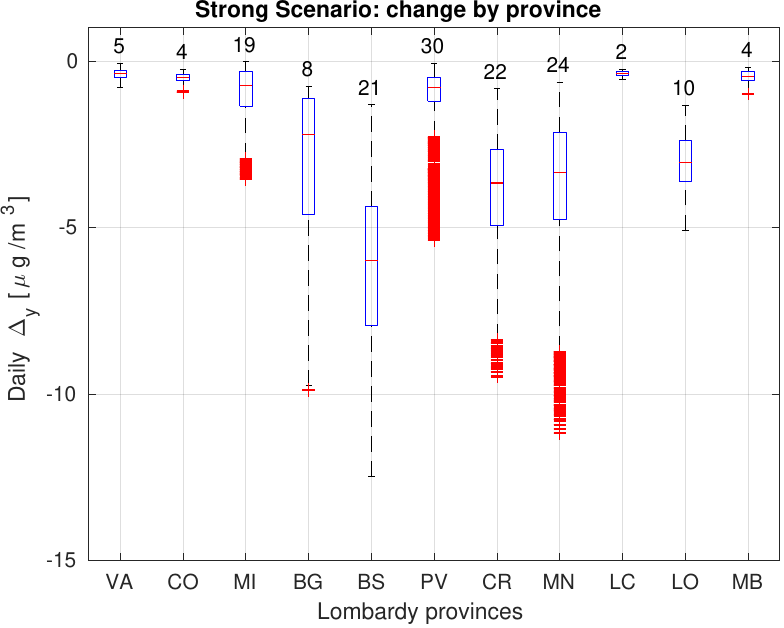}
        }%
    \qquad
    \subfloat[][]{   
        \label{fig:Scenario_B_land_boxplot}
        \includegraphics[scale=0.57]{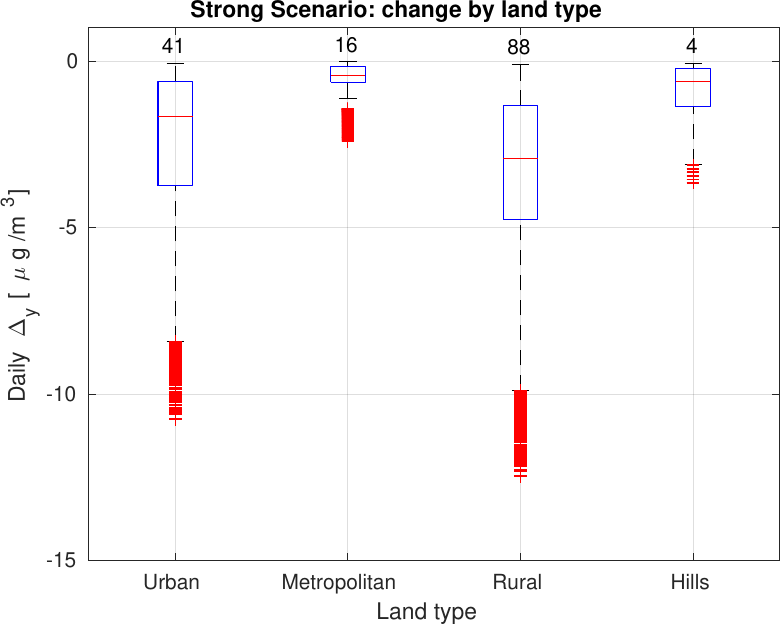}
        }%
    \caption{Scenario Strong (-50\%). Boxplot of daily $\Delta_{\hat{y}}$ in winter (452 daily observations for each pixel). Panel a: grouped by province; panel b: grouped by land type. The number of pixels of each category is reported over the corresponding boxplot. The model is evaluated only in the non-forested area under 640 $m$ altitude.}%
    \label{fig:Scenario_B_boxplot}%
\end{figure}

Furthermore, Figure \ref{fig:Scenario_A_boxplot} and \ref{fig:Scenario_B_boxplot} show the distribution of the winter daily PM$_{2.5}$ reduction aggregated by province and land use for the two scenarios considered. As expected, the largest reductions are obtained in Brescia, Cremona and Mantua provinces, which are characterised by large rural areas and extensive livestock farming. It is also shown that metropolitan and hill areas are not affected by an important reduction.

Table \ref{tab:scenario_comparison} summarises the average of PM$_{2.5}$ reductions aggregated by province in $\mu g/m^3$.  Note that, due to the presence of the Prealps and Alps, the scenario analysis covers a limited surface of the Lecco, Varese, and Como provinces. Finally, Table \ref{tab:scenario_comparison_season} shows the change aggregated by season. As expected, the largest effect is in winter, when NH$_3$ plays a limiting role in the formation of PM$_{2.5}$. We observe that in spring we have no effect as the increase is non-significant being smaller than $0.1$\% and with a $t$-statistics not larger than $0.125$.  In winter, when the NH$_3$ emissions are less abundant, the overall estimated reduction over the plain areas for the PRIA scenario is $1.44$ ($\text{std} = 0.08$) while for the Strong scenario is $2.76$ ($\text{std} = 0.16$). These results can be compared with the PM$_{2.5}$ estimated average $39.49$ $\mu g/m^3$. So, considering the PRIA scenario, the PM$_{2.5}$ overall reduction is close to 3.5\% while considering the Strong scenario, the PM$_{2.5}$ overall reduction is close to 7\%.

\begin{table}[h]
    \centering
\caption{Winter change by province and scenario. $\mathcal{D}^{*}$ is the number of pixels involved in the computation, $\bar{y}$ is the estimated average of the PM$_{2.5}$ concentrations; the columns headed PRIA and Strong show the average reduction and its uncertainty computed using Eq. (\ref{eq:aggregation_delta_var}) for both scenarios.}
\label{tab:scenario_comparison}
\begin{tabular}{lrc|c|c}
\toprule
& & $\bar{y}$ & PRIA & Strong \\
Province & $\mathcal{D}^{*}$  & $\mu g/m^3$     & $\mu g/m^3$  & $\mu g/m^3$   \\
\midrule
VA & 5   & 32.78& -0.19 (0.09)& -0.37 (0.09)\\
CO & 4   & 36.26& -0.26 (0.11)& -0.50 (0.11)\\
MI & 19  & 37.56& -0.47 (0.03)& -0.90 (0.06)\\
BG & 8   & 40.60& -1.62 (0.11)& -3.11 (0.19)\\
BS & 21  & 47.13& -3.14 (0.18)& -6.04 (0.34)\\
PV & 30  & 36.10& -0.50 (0.03)& -0.96 (0.06)\\
CR & 22  & 41.33& -2.04 (0.12)& -3.92 (0.22)\\
MN & 24  & 39.64& -1.97 (0.11)& -3.80 (0.22)\\
LC & 2   & 34.50& -0.19 (0.21)& -0.37 (0.21)\\
LO & 10  & 38.41& -1.57 (0.10)& -3.02 (0.18)\\
MB & 4   & 37.57& -0.24 (0.11)& -0.46 (0.11)\\
Overall & 149 & 39.49& -1.44 (0.08)& -2.76 (0.16) \\
\bottomrule
\end{tabular}
\end{table}

\begin{table}[h]
    \centering
\caption{PM$_{2.5}$ average change, $\bar{\Delta}_{\hat{y}}(\bm{s}, t)$, by season and scenario. The number of pixels involved in the computation is $\mathcal{D}^{*} = 149$. $\bar{y}$ is the estimated average of the PM$_{2.5}$ concentrations; the columns headed PRIA and Strong show the average reduction and its uncertainty computed using Eq.  (\ref{eq:aggregation_delta_var}) and the (\%) of reductions for both scenarios. }
\label{tab:scenario_comparison_season}
\begin{tabular}{m{0.8cm}c|rr|rr}
\toprule
& $\bar{y}$ & \multicolumn{2}{c}{PRIA} & \multicolumn{2}{c}{Strong} \\
Season & $\mu g/m^3$     & $\mu g/m^3$ & \%  & $\mu g/m^3$ & \%   \\
\midrule
Autumn & 24.49&  -0.66 (0.07)& -2.68 & -1.26 (0.14)& -5.15\\
Spring & 18.57&  0.01 (0.08) & 0.03  & 0.01 (0.15) &  0.07\\
Summer & 13.92&  -0.25 (0.08)& -1.81 & -0.49 (0.15)& -3.48\\
Winter & 39.49&  -1.44 (0.08)& -3.64 & -2.76 (0.16)& -7.00\\
Overall & 24.05&-0.58 (0.04)& -2.42& -1.12 (0.08)& -4.64\\
\bottomrule
\end{tabular}
\end{table}

\section{Discussion}
\label{sec:Discussion}
This study quantified the reduction in PM$_{2.5}$ concentrations achieved through a reduction in NH$_3$ emissions for the Lombardy Plain. If we look at the entire area considered and the entire period, against an ammonia emissions reduction of 26\% (resulting from the application of the PRIA regional air quality plan), we obtain a significant fine particulate reduction close to 2.4\%. A reduction of 50\% in ammonia emissions improves the air quality by 4.6\%. On the other hand, the absolute reduction in PM$_{2.5}$ depends on the fraction of fine-particulate mass that is directly ammonia-sensitive. In the Lombardy Plain, the composition of PM$_{2.5}$ in winter has a very important secondary component, usually exceeding 50\%. As a result, the greatest reductions are obtained in winter (over 3.6\% and 7\% for the two scenarios, respectively). 

Looking at the spatial distribution of the reductions, it can be seen that the most benefited areas are in the southeastern area of the region, those areas corresponding to the high density of livestock farms and, therefore, ammonia emissions. Consequently, Lombardy has the largest potential of reducing winter averaged PM$_{2.5}$, considered beneficial to human health, by strongly controlling NH$_3$ emissions. 

\section{Conclusions and further developments}
\label{sec:Conclusions}
The results of this study are generally consistent with those obtained using chemical transport models. Considering the root mean square error, our results are, in some cases, better than \cite{veratti2023impact}. This means that a detailed statistical model fitted to an extensive dataset may catch the main features of a chemical transport model. 
Since the computational burden is definitely lower, these results hint at the use of statistical emulators for policy impact assessment.

Our results are also consistent with those in \cite{fasso2023extent}, which are based on observed ammonia concentrations instead of inventory emissions. This means that the emission data are reliable for understanding the impact of livestock on air quality.

From the methodological point of view, our proposal, considering an unstructured skedastic function, may be considered as a first step, opening the development of deterministic and/or stochastic skedastic functions characterised by a smaller number of degrees of freedom. The software for implementing the heteroskedastic HDGM developed in this study represents an updated version of the open-source D-STEM software \cite{FinazziDSTEMv2}. It can be accessed on the GitHub repository at \url{https://github.com/graspa-group/d-stem}.

The monitoring network spatial unbalance mentioned in the introduction and mapped in Figure \ref{fig:emission_all} means that our model has not been trained where NH$_3$ emissions are very high.
For this reason, we consider our results cautionary and think that the impact computed may underestimate the true impact. Further research is needed to understand this. Since the Lombardy sampling bias is a consequence of the European Union (EU) rules not requiring NH$_3$, it is not easy to fill this gap for the EU. One possibility is to validate our approach using Swiss data \cite{GRANGE2023}, which has good temporal and spatial coverage.

\section*{Credit author statement}
Alessandro Fassò: Conceptualisation, Methodology, Supervision. Jacopo Rodeschini: Methodology, Formal analysis, Software, Data curation, Writing original draft. Francesco Finazzi: Methodology, Revision of the article. Alessandro Fusta Moro: Revision of the article.  

\section*{Acknowledgements}
This research was co-funded by Fondazione Cariplo under the grant 2020–4066 \textit{“AgrImOnIA: the
impact of agriculture on air quality and the COVID-19 pandemic”} from the \textit{“Data Science for
Science and Society”} program and by the European Union -  NextGenerationEU, in the framework of the \textit{“GRINS - Growing Resilient, INclusive and Sustainable”} project (GRINS PE00000018 – CUP F83C22001720001). The views and opinions expressed are solely those of the authors and do not necessarily reflect those of the European Union, nor can the European Union be held responsible for them.

\bibliographystyle{abbrvnat}
\bibliography{ref}



\end{document}